\documentclass[fleqn,12pt]{wlscirep}
\usepackage[utf8]{inputenc}
\usepackage[T1]{fontenc}

\usepackage{amsmath}
\usepackage{amssymb}
\usepackage{graphicx}
\usepackage[subrefformat=parens]{subcaption}

\usepackage{lineno}

\makeatletter
\let\LN@align\align
\let\LN@endalign\endalign
\renewcommand{\align}{\linenomath\LN@align}
\renewcommand{\endalign}{\LN@endalign\endlinenomath}
\let\LN@gather\gather
\let\LN@endgather\endgather
\renewcommand{\gather}{\linenomath\LN@gather}
\renewcommand{\endgather}{\LN@endgather\endlinenomath}
\let\LN@abstract\abstract
\let\LN@endabstract\endabstract
\renewcommand{\abstract}{\linenomath\LN@abstract}
\renewcommand{\endabstract}{\LN@endabstract\endlinenomath}
\makeatother

\title{Collective patterns and stable misunderstandings in networks striving for consensus without a common value system }

\author[1,*]{Johannes Falk}
\author[2,3]{Edwin Eichler}
\author[3,4]{Katja Windt}
\author[1]{Marc-Thorsten Hütt}
\affil[1]{Department of Life Sciences and Chemistry, Jacobs University, Bremen, Germany}
\affil[2]{EICHLER Consulting AG, Weggis, Switzerland}
\affil[3]{SMS Group GmbH, Düsseldorf, Germany}
\affil[4]{Global Production Logistics, Jacobs University Bremen, Germany}

\affil[*]{j.falk@jacobs-university.de}

\date{}

\begin{abstract}
Collective phenomena in systems of interacting agents have helped us understand diverse social, ecological and biological observations. The corresponding explanations are challenged by incorrect information processing. In particular, the models typically assume a shared understanding of signals or a common truth or value system, i.e., an agreement of whether the measurement or perception of information is `right' or `wrong'. It is an open question whether a collective consensus can emerge without these conditions
 
Here we introduce a model of interacting agents that strive for consensus, however, each with only a subjective perception of the world. Our communication model does not presuppose a definition of right or wrong and the actors can hence not distinguish between correct and incorrect observations. Depending on a single parameter that governs how responsive the agents are to changing their world-view we observe a transition between an unordered phase of individuals that are not able to communicate with each other and a phase of an emerging shared signalling framework. 

We find that there are two types of convention-aligned clusters: one, where all social actors in the cluster have the same set of conventions, and one, where neighbouring actors have different but compatible conventions (`stable misunderstandings').
\end{abstract}

\begin{document}






\flushbottom
\maketitle

\thispagestyle{empty}


\section{Introduction}

Self-organisation in networks of communicating agents is a fundamental process for the emergence of order in social systems~\cite{kearnsExperimentalStudyColoring2006a,kearnsBehavioralExperimentsBiased2009b,helbingSelfOrganizationEmergenceSocial2011}. Such spontaneous order occurs in the form of consensus~\cite{gavriletsConvergenceConsensusHeterogeneous2016}, social segregation and opinion formation~\cite{flaxmanFilterBubblesEcho2016,masIndividualizationDrivingForce2010} or as the establishment of norms and conventions, e.g. a language or set of signals~\cite{ehrlichEvolutionNorms2005,hawkinsEmergenceSocialNorms2019,lipowskaEmergenceLinguisticConventions2018,barrEstablishingConventionalCommunication2004a,spikeMinimalRequirementsEmergence2017a}. Beyond the social sciences, also 
Biology~\cite{oliphantLearningEmergenceCoordinated, erezCelltocellInformationFeedbackinduced2020} and Computer Science~\cite{grolikInformationLogisticsDecentralized2012, anthonyEmergenceParadigmRobust2004, kirbyNaturalLanguageArtificial2002, staabEmergentSemantics2002, steelsEvolvingGroundedCommunication2003} are interested in how interacting agents do and could organise.

Consequently, there exist a variety of models that  analyse or explain different aspects of self-organisation within communities: In the Voter Model~\cite{cliffordModelSpatialConflict1973a, holleyErgodicTheoremsWeakly1975b, rednerRealityinspiredVoterModels2019} or the Axelrod Model~\cite{axelrodDisseminationCultureModel1997}, spatially distributed agents copy or invert opinions or attributes from their neighbours according to certain rules, globally leading to a transition between an ordered homogeneous and a disordered state~\cite{shaoDynamicOpinionModel2009, gambaroInfluenceContrariansDynamics2017, klemmNonequilibriumTransitionsComplex2003, nyczkaPhaseTransitionsVoter2012}. The Lewis Signalling Game~\cite{lewisConventionPhilosophicalStudy2002} explains how a sender and a receiver agree on common signals and the naming game~\cite{steelsSelfOrganizingSpatialVocabulary1995,baronchelliGentleIntroductionMinimal2016} describes how agents agree on a common vocabulary for objects in their environment. Another approach is the analysis of learning in social networks~\cite{golubLearningSocialNetworks2016}. As an example, DeGroot~\cite{degrootReachingConsensus1974} showed how agents that start with individual (subjective) knowledge about a parameter can reach a consensus by repeatedly assimilating information about the parameter that they observe in the rest of the group. Besides these purely theoretical studies, there are also findings with empirical support. Building a dynamical model around given data of scientific collaboration it was e.g. possible to find signatures of self-organisation in real social processes~\cite{dankulovDynamicsMeaningfulSocial2015}.

Less investigated is the emergence of order (or synchronisation) under erroneous or subjective perception and in the absence of a common value system, an objective instance or a universal truth. Exceptions to the first point (erroneous perception) include the effect of information processing noise on vote dynamics~\cite{moreiraEfficientSystemwideCoordination2004a, granovskyNoisyVoterModel1995} or naming games~\cite{limNoisyNamingGames2011}. The lack of a common value system has partly been the motivation behind variations of the naming
game~\cite{baronchelliNonequilibriumPhaseTransition2007,steelsSelfOrganizingSpatialVocabulary1995,baronchelliGentleIntroductionMinimal2016}, the signalling game~\cite{skyrmsEvolutionSignallingSystems2009,skyrmsDynamicModelSocial2000, hutteggerDynamicsSignalingGames2014, zollmanTalkingNeighborsEvolution2005, barrEstablishingConventionalCommunication2004a,baronchelliTopologyinducedCoarseningLanguage2006,baronchelliGentleIntroductionMinimal2016, louLocalCommunitiesObstruct2018} and several other models~\cite{caveLearningAgree1983, parikhCommunicationConsensusKnowledge1990, heifetzCommentConsensusCommon1996},  even though all these models still rely on the global notion of `true' and `false' or a shared interpretation of signals/actions.

Here, we present a model that addresses both challenges -- erroneous perceptions and lack of a universal truth -- simultaneously.  Our model comprises a network of agents that each only has a subjective perception of the world. Thereby, they are faced with a cognitive dissonance between their own cognition and what they perceive as their neighbours' cognition~\cite{groeberDissonanceMinimizationMicrofoundation2014}. In order to reach conformity~\cite{cialdiniSocialInfluenceCompliance2004}, each agent strives to minimise its cognitive dissonance, based only on its own subjective observations. We show that under these conditions collective behaviour still can emerge. We also show that stable misunderstandings can form, i.e. an emerging pattern where many nodes are without perceived conflict, despite heterogeneity in subjective perception. Often this emerges as an alternating arrangement of compatible but distinct subjective perceptions on suitable network topologies.
To elaborate on the interplay between our model's dynamics and the topology of the underlying network, we analyse the dynamics on regular lattices and random regular graphs. The results help to better understand the emergence of order within a connected community of agents without an objective instance.

Our model is motivated by Gotthard Günther's polycontextural logic~\cite{guntherIdeeUndGrundriss1991, guentherBeitraegeZurGrundlegung1976}, where two subjects observing the same situation can come to different conclusions, even when each of the subjects adheres to binary (but distinct, contexture-dependent) logic. We hence denote our model \textit{polycontextural networks}.

Our findings not only have an impact on sociological questions but also on the ongoing problem of how distributed machine learning systems can negotiate a common signalling system. It is also an ongoing debate in philosophy how a consensus can emerge out of observer-dependent facts~\cite{dibiagioStableFactsRelative2021}.

The remainder of the paper is organised as follows: In the next section, we introduce our model and illustrate its static properties with two simple network motifs. In Sec.~\ref{sec:result} we then analyse the dynamics of the model on random regular graphs, before we focus on triangular and square lattices to investigate the observed self-similarity and develop a mechanistic understanding of the observed dynamics. Subsequently, in Sec.~\ref{sec:discussion} we discuss the implications of our model and draw some conclusions in Sec.~\ref{sec:conclusion}.

\section{Model}
\label{sec:model}
The polycontextural network is a simple model where $N$ agents interact over a network. Each agent $A_n$ with $n \in N$ is equipped with a characteristic $c_n$ whose expression is taken from a pool of size $C$. To simplify the notation, the characteristic of each agent is given as a standard basis vector $\boldsymbol{e}_i$ of length $C$ with $1$ in the $i$th position and 0 in every other position.

\begin{figure}
    \centering
    \includegraphics[width=.5\textwidth]{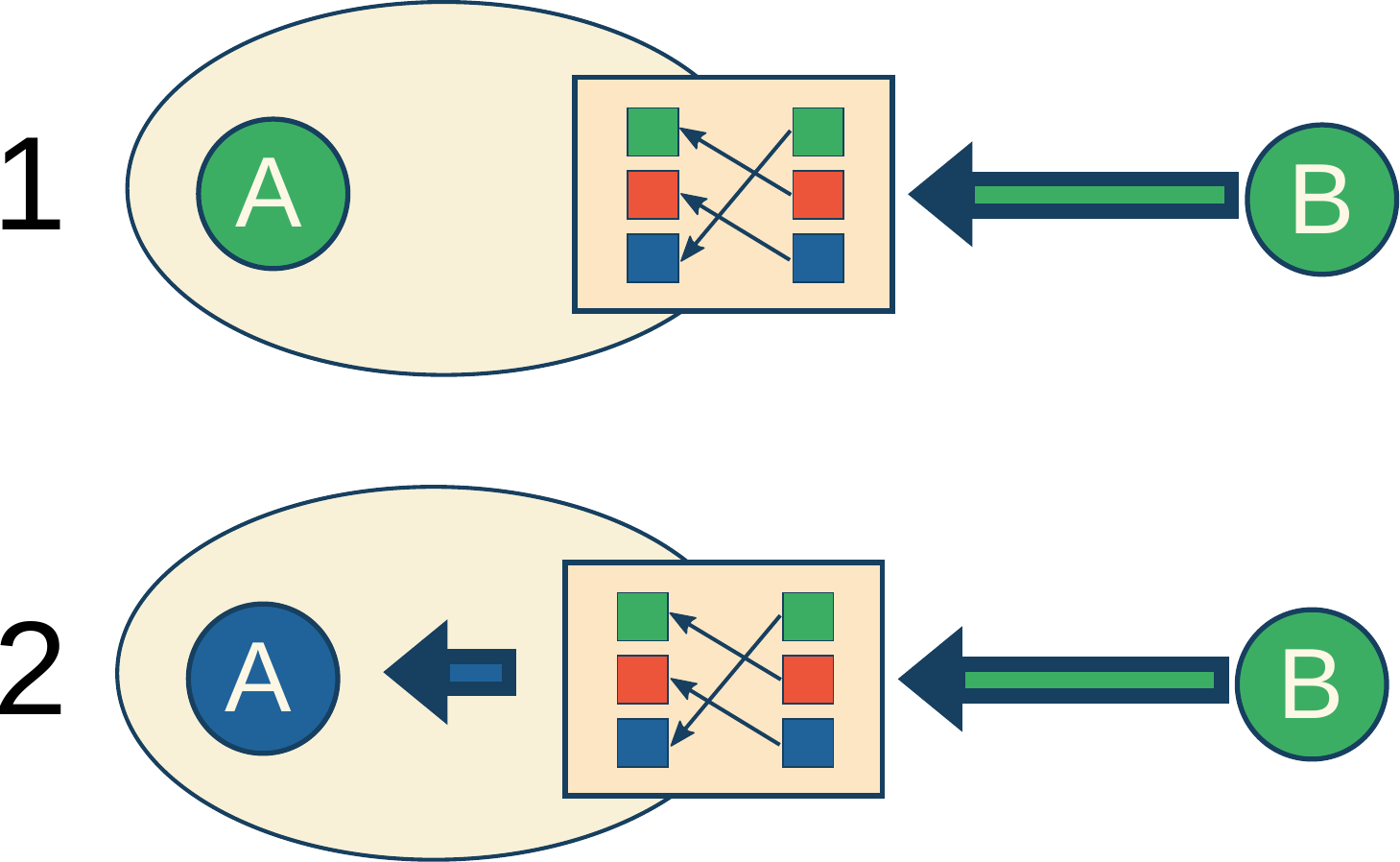}
    \caption{(top) $A$ observes the colour of agent $B$. From an objective view, $B$'s colour is green. \\(bottom) However, due to $A$'s dictionary, $A$ observes blue and changes its colour accordingly.}
    \label{fig:tables1}
\end{figure}

To incorporate subjectivity, each agent has an individual dictionary that bijectively maps the `outside world' of the agent to its personal cognition. Formally, this dictionary is a bijective function $\sigma: C \to C$ and can be written as a $C \times C$ permutation matrix $T_n$. If one agent $A_n$ observes the characteristic $c_m$ of another agent, the observing agent sees $T_n c_m$ instead of the `true' (objective) $c_m$ (as depicted in Fig.~\ref{fig:tables1}). To give it an intuitive meaning, in the following we will assume that the characteristics $c_n$ are colours. Due to this definition, our model does not have objective truth values -- a predefined understanding of colour -- but $C!$ different and equally correct world-views (here: colour mappings). In the following, we understand the term \textit{world-view} to mean a set of truth values that determine how an agent perceives the environment. Each world-view hence refers to a specific choice of a value system.

The different agents are spatially distributed and partially connected, whereby they form a network structure where each agent is one node. During this investigation, we will analyse our model with different network topologies of different sizes. 

The only interaction in our model is a simple version of social influence where all agents strive for consensus.
A single update step of our model proceeds similarly to the standard voter model with $C$ different opinions and asynchronous dynamics, which means that a randomly chosen agent $A_n$ adopts the opinion (the colour) of one of its neighbours $A_m$. However, and in difference to the voter model, $A_n$ can not observe the `true and objective' colour $c_m$ but sees the characteristic $T_n c_m$. A sequence of $N$ updates forms a time step, which means that on average at every time step every node is selected once.

While formally similar to the voter model, due to the different world-views ($T_x$) and hence the different perception of colours, the model's dynamic would in general not converge to a uniform colouring. To illustrate this, let us imagine two nodes A and B, which can be either red or blue. The nodes shall have two different, but fixed world-views: A recognises colours as they are and B recognises the colours reversed (blue to red, red to blue). If A is red and is observed by B, B will turn blue. In some subsequent time step, A will observe B and also turn blue. However, as soon as B observes A again, B will turn red and the process starts again.

What is missing is the single agent's ability to sense that its own world-view is not aligned with that of its neighbours. According to relational epistemology~\cite{thayer-baconNurturingRelationalEpistemology1997} and the philosophy of world-views~\cite{sireNamingElephantWorldview2015,klubackDiltheyPhilosophyExistence1957}, world-views are shaped by and changed according to lived experiences and determine how one understands the world and responds to it.

Following this concept, within our model each agent is equipped with two internal counters: $O_n$ and $K_n$, and every update step proceeds as follows: 
\begin{itemize}
\item One agent $A_n$ and one of its neighbours $A_m$ are selected randomly following a uniform distribution. 
\item Agent $A_n$ subjectively observes the feature of $A_m$, which means $A_n$ sees the characteristic $T_n c_m$. 
\item If agent $A_n$'s own characteristic is already equal to the observed one, $A_n$ only increments its internal $O_n$ counter by 1.
\item Otherwise, $A_n$ changes its own characteristic to the subjectively observed one and increments both its internal $O_n$ and $K_n$ counter by one.
\item If the fraction $K_n$/$O_n$ is larger than the parameter $q$ (which means that in more than $q$ percent of the observations the observed characteristic was not equal to the own), the agent changes its own world-view $T_n$ to a random selection out of the $C!$ possibilities and resets both counters to zero. Note that $O_n$ gets always incremented at least once before this last step. The fraction is hence always defined.
\end{itemize}

The individuals in our model hence share a predefined response once a threshold number of conflicts is detected and, following the definitions given in Ref~\cite{baronchelliEmergenceConsensusPrimer2018}, our model would have to be considered to belong to the group of quorum sensing models, although the term `subjective quorum-sensing' would probably fit best. In terms of everyday experience, it may seem strange that opinions in our model are changed immediately, regardless of past observations. However, what our model reflects are the different time scales for a change of opinion vs. a change of world-view.

The illustrative idea behind the dynamics of the model is that agents react -- according to their subjective interpretation -- to the states of other agents. Since each agent is also an object of other observations, the state of the observed neighbours is sometimes already the reaction to the observation of the own state. This enables each agent to perform a self-reflection and a repeated observation of the neighbour's state can hence indicate whether the neighbourhood confirms the own world-view (for a more detailed philosophical interpretation see Sec.~\ref{sec:discussion}). The parameter $q$ (threshold parameter) therefore controls how sensitive an agent decides that its own belief system does not conform to the neighbourhood, subsequently changing it.

Before we proceed, we define two special terms to simplify the notation:
Two connected nodes $i,j$ are called \textit{compatible}, if their dictionaries mutually agree in all colours, which means $T_i \times T_j = \mathbb{I}$. A network is considered \textit{solved} if all connected nodes are compatible. Note that our definition of compatible and solved networks only depends on the dictionary of the nodes and not on the current colour. As we will show in the following examples, a solved network does not imply that no more colour changes occur. 

\subsection{Two illustrative examples}

\begin{figure}
    \centering
    \includegraphics[width=.5\textwidth]{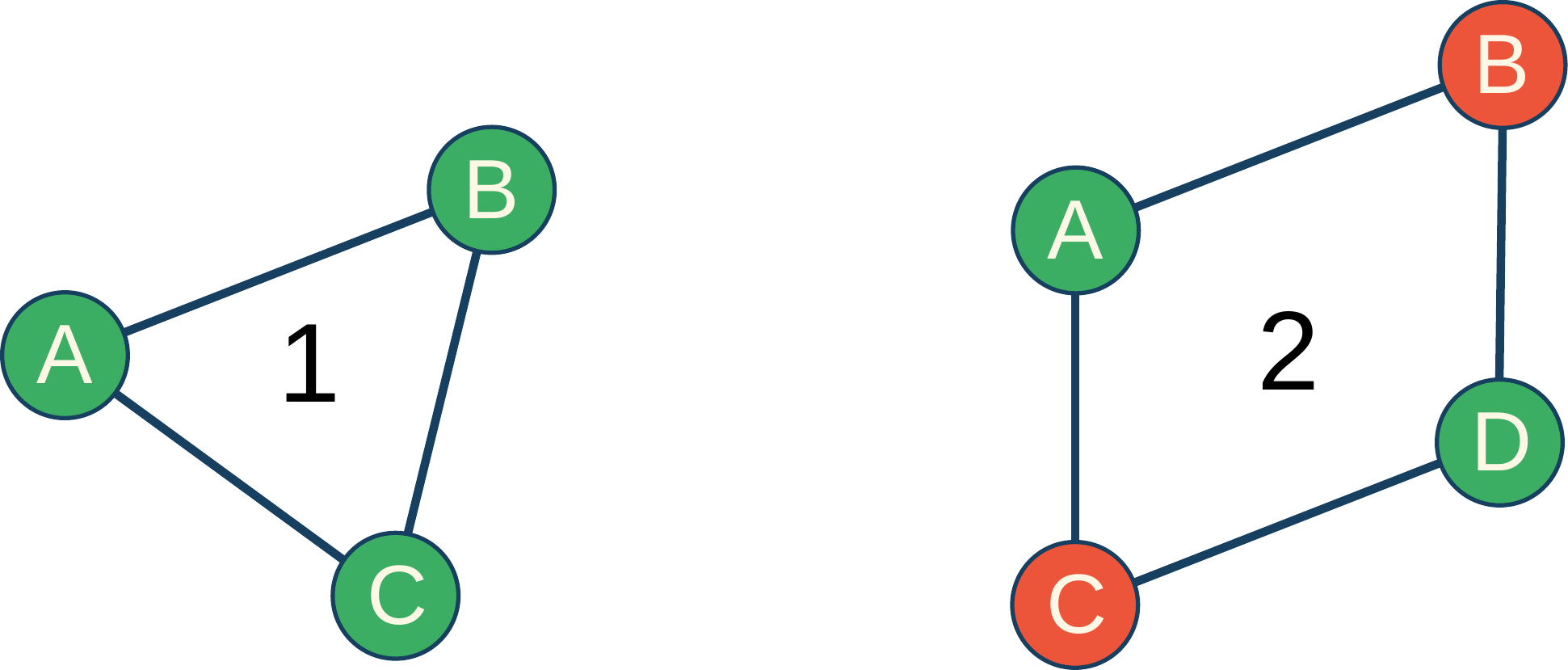}
    \caption{Two different minimal models to illustrate possible solved states of our model.}
    \label{fig:netmotif}
\end{figure}

Let us refer to two minimal network motifs as illustrative examples. Fig.~\ref{fig:netmotif}(1) shows three connected nodes (a triangular structure). Following the above-given definition of a solved system, node $A$ has to be compatible with respect to both nodes $B$ and $C$, which means:
\begin{align}
    T_A \times T_B &= \mathbb{I} \\
    T_A \times T_C &= \mathbb{I}.
\end{align}
At the same time, the nodes $B$ and $C$ need to be compatible as well, hence:
\begin{align}
    T_B \times T_C = \mathbb{I}.
\end{align}
This is only possible if
\begin{align}
    T_A = T_B = T_C = T_A^{-1},
\end{align}
which is true for all 2-cycle permutation matrices.
Note that the definition of a solved system is a local definition and does explicitly not imply that no more colour changes occur. Let us hereto assume that $C=3$ and for all dictionaries it holds:
\begin{align}
   T_{\{A,B,C\}} = \begin{pmatrix}
    0 & 1 & 0 \\
    1 & 0 & 0 \\
    0 & 0 & 1 
    \end{pmatrix}.
    \end{align}
Let us further assume that node $A$ is in state $c_A = (1,0,0)$. If both node $B$ and $C$ observe $A$, both change their states to $c_{A/B}=(0,1,0)$. However, if now $B$ observes $C$ it needs to change its state again to $c_B=(1,0,0)$ which would create a further state change after an observation by $A$, and so on. For the system to reach a state without any more colour changes, it needs to hold:
\begin{align}
    T_A T_B T_C = T_A^3 = \mathbb{I}.
\end{align}
This would require $T_A$ to be a 3-cycle permutation matrix. However, since we already know that $T_A$ needs to be a 2-cycle permutation matrix this is only possible if $T_A$ is the identity. 

Things are different for the four-node system shown in Fig.~\ref{fig:netmotif}(2). For $A$ it again holds:
\begin{align}
    T_A \times T_B &= \mathbb{I} \\
    T_A \times T_C &= \mathbb{I}.
\end{align}
However, $B$ and $C$ are not connected directly, but only via $D$. To reach a solved system there are hence two more equations that need to be satisfied:
\begin{align}
    T_D \times T_B &= \mathbb{I} \\
    T_D \times T_C &= \mathbb{I},
\end{align}
which means that:
\begin{align}
    T_A &= T_D = T_B^{-1}\\
    T_B &= T_C = T_A^{-1}.
\end{align}
In contrast to the triangular structure, these constraints cause a system where no more changes of colour will happen. To see how this comes, let us start with node $A$ and cycle over the other nodes. It follows:
\begin{align}
    T_B T_D T_C T_A = T_B T_A T_B T_A = T_B T_B^{-1} T_B T_B^{-1} = \mathbb{I}.
\end{align}
Since in this configuration a system can be stable although the agents have different world-views, we call this novel effect a \textit{stable misunderstanding}. The important point here is that -- due to their subjectivity -- the nodes involved can not detect this misunderstanding. To our knowledge, our polycontextural network model is the first that allows for and demonstrates the impact of such stable misunderstandings.

One might argue that this observed effect of stable misunderstandings is just an artefact of whether the considered cycle has an even or odd number of nodes. While this is true from a mathematical point of view, we argue that the artefact mainly arises because of our simplified world-view representation. We assume that for each node perception of colours takes place deterministically (no single color is perceived in multiple ways without a change in world-view) and without information loss (two distinct colours are never perceived as (the) same colour; no `colour blindness'). This set of requirements leads to the restriction to bijective translation tables. If we would e.g. allow for non-bijective associations within the translation tables, the effect would vanish. However, to keep the model simple and comprehensible, we will stick to our definition.

Over time, we observe that connected nodes synchronise their world-views and form clusters of nodes with the same understanding of the world. We are hence mainly interested in the dynamics and organisation of the world-views (the tables). The (fluctuating) colours of the nodes are just signs (their language) to communicate with their surroundings and are -- in our investigation -- only of limited interest. In the following section we show that, depending on the value of the threshold value, the sizes of these table-clusters stay small or expand over all scales, indicating a critical state and a phase transition. Besides the change of the threshold parameter, we also demonstrate how the topology of the network affects the type and size of clusters.

In what follows, we present the results of the simulations for four different network topologies. To avoid any grid artefacts, we first analyse random 3-regular and 4-regular graphs. To better understand and visualise the dynamics of our model we then focus on regular triangular and square lattices.  For all models, we set $C=3$ and -- to avoid boundary effects -- use periodic boundary conditions if applicable.

\section{Results}
\label{sec:result}

\subsection{Phenomenology}

\begin{figure}
    \centering
    \includegraphics[width=.8\textwidth]{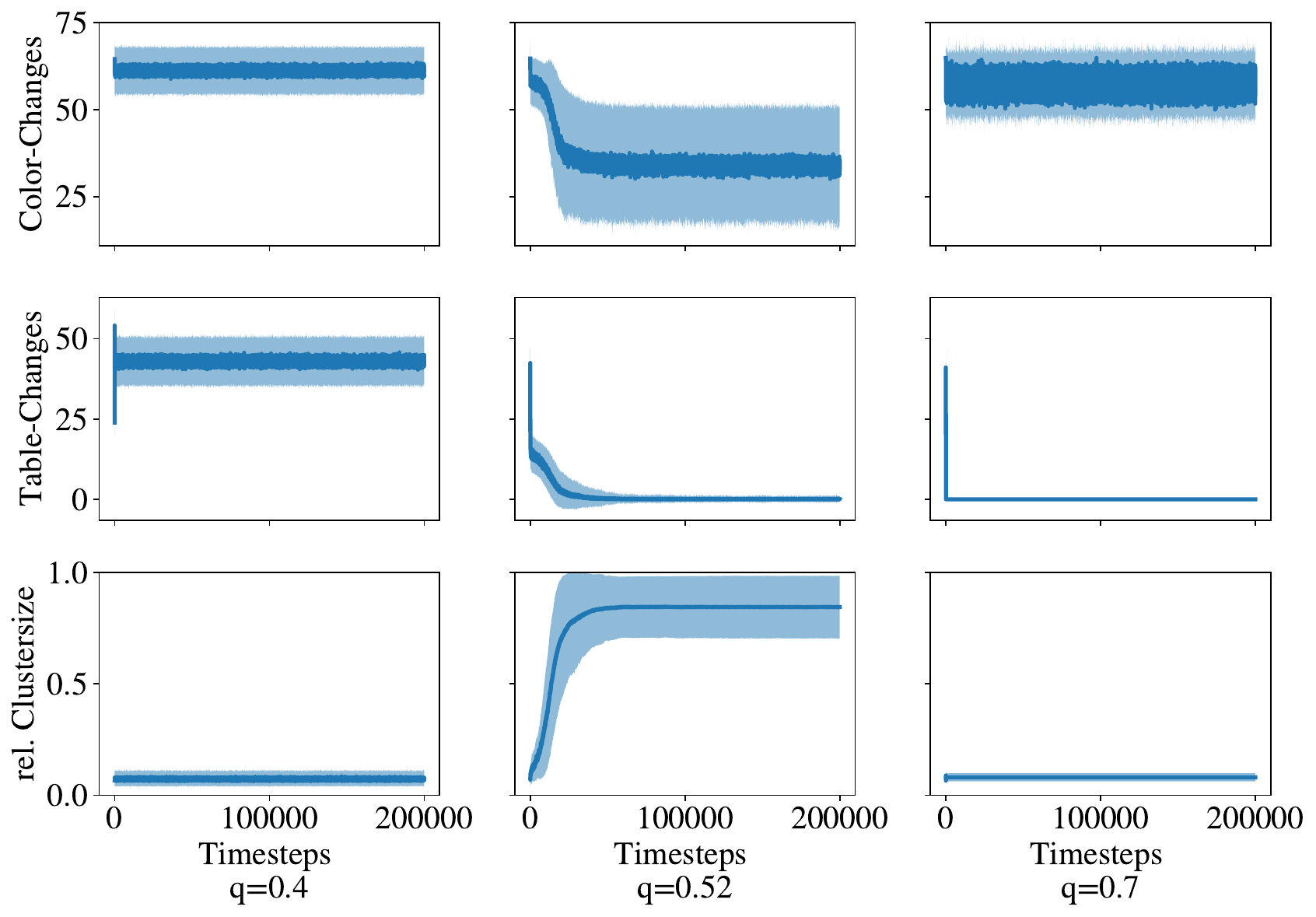}
    \caption{Time series of colour changes per time step, table changes per time step and relative mean size of the clusters for different values of the parameter $q$ (average over $100$ runs). The light blue area indicates the standard deviation. The analysed network is a random 4-regular graph with $200$ nodes.}
    \label{fig:time_rr}
\end{figure}

We will first illustrate the time-dependent behaviour of our model before we proceed to analyse the dependence on the $q$ parameter.
Fig.~\ref{fig:time_rr} shows the time evolution of colour changes per time step, table changes per time step and the relative mean size of clusters of the same translation table $T_X$ within a random 4-regular lattice with 200 nodes for three different values for $q$. If the threshold value is small ($q < q_c$) the nodes constantly change their world-views, corresponding to a constant number of table changes. These frequent changes prevent a build-up of clusters with equal world-views. Since no clusters of compatible world-views emerge, there are also frequent colour changes. If the threshold value is too large $q > q_c$, there are only a few table changes before the system reaches a frozen state (the agents are satisfied with their world-view). The dynamics freeze before large clusters can emerge, hence neighbouring world-views are incompatible and frequent colour changes are observed. For an intermediate value $q \approx q_c$ we observe a slow decrease in the change of the world-view tables that finally leads to the build-up of a large cluster, similar to the behaviour observed in naming games~\cite{baronchelliSharpTransitionShared2006a} or in threshold voter models~\cite{ehrlichEvolutionNorms2005}. 

\begin{figure}
    \centering
    \includegraphics[width=.6\textwidth]{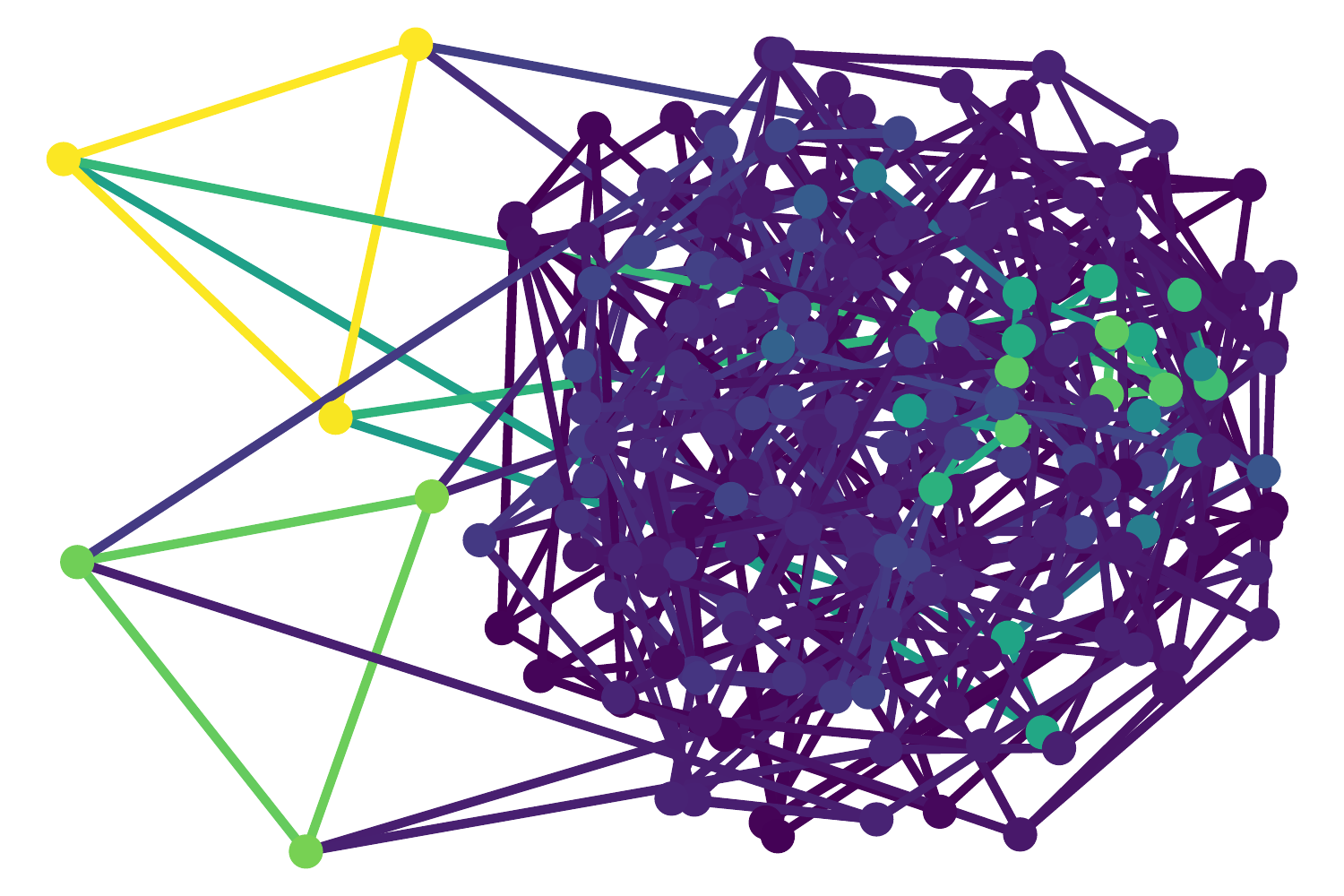}
    \caption{Illustration of a random 4-regular graph. The colours indicated the time since the last change of the respective node. The more yellowish the node, the longer no change of the translation table. The six nodes with the longest time since the last change have been shifted out of the bulk for illustrative purposes.}
    \label{fig:rr_top}
\end{figure}

To gain insight into how the growth of clusters is taking place, Fig.~\ref{fig:rr_top} shows a snapshot of a random 4-regular graph. The colours of the nodes indicate the time span the respective node has not changed. The more yellowish the node, the longer no change. For illustrative purposes, the six nodes that did not change for the longest have been shifted out of the bulk. As can be seen from the figure, three of these nodes each form a separate triangular motif. These triangular motifs were the nuclei for the growth of a large cluster of shared world-views. The evolution of the cluster sizes can thus be understood as a nucleation and coarsening process. If the cluster does not span the full system we can still see colour fluctuations, however, the colours change less often than in the other two cases. There is hence an intermediate regime between continuous table changes and an immediately frozen state, where world-views can synchronise and cluster.

\subsection{q-Dependence}

\begin{figure}
    \centering
    \includegraphics[width=.9\textwidth]{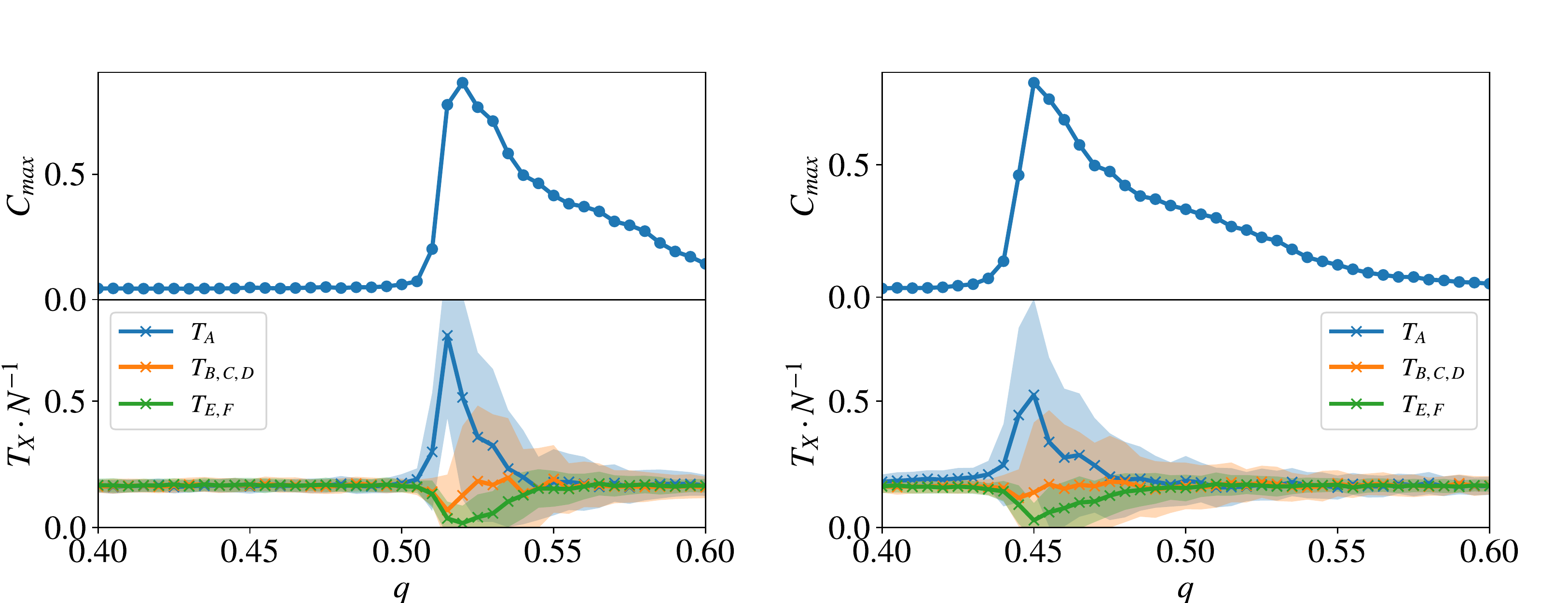}
    \caption{(upper) Relative cluster size of the largest cluster $C_{max}$ vs. the threshold parameter $q$ on an (left)~random 3-regular graph and (right)~random 4-regular graph ($N = 200$, after $t=80,000$ simulated steps, averaged over $100$ runs). (lower) The fraction of the different tables depending on $q$. Here, $T_X$ denotes the different types of possible translation tables as described in the full text. The light area around the curves indicates the standard deviation.}
    \label{fig:rr_grid_tables}
\end{figure}

To gain more insight into the critical behaviour, in the following we analyse in detail how the model behaves under a change of the parameter $q$.
Figure~\ref{fig:rr_grid_tables} shows exemplary behaviours of the relative size of the largest cluster over the value of $q$ for both a random 3-regular and 4-regular graph of size $N=200$. For a small window of $q$ the largest cluster spans the full system, indicating system-wide correlations between the tables. Additionally, the lower plots show the fraction of the six different translation tables, where $T_A$ denotes the identity, $T_{\{B,C,D\}}$ denote the three possible 2-cycle matrices and $T_{\{E,F\}}$ denote the two 3-cycle matrices, with $T_E \times T_F = \mathbb{I}$.

In terms of the motivation of our model the results so far already prove the emergence of a consensus. However, to better understand the characteristics of our model as well as its properties at criticality, we will now turn to two simplified network topologies and analyse their behaviour.

\subsection{Regular Lattices}
\begin{figure}
\centering
\begin{subfigure}[c]{0.3\textwidth}
    \centering
    \includegraphics[width=.9\textwidth]{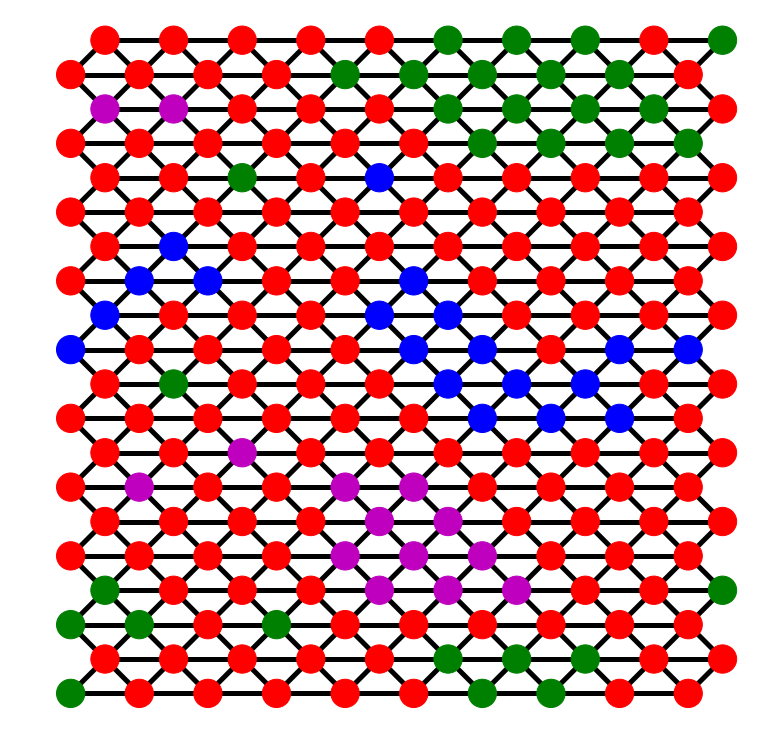}
\end{subfigure}
\begin{subfigure}[c]{0.3\textwidth}
    \centering
    \includegraphics[width=.9\textwidth]{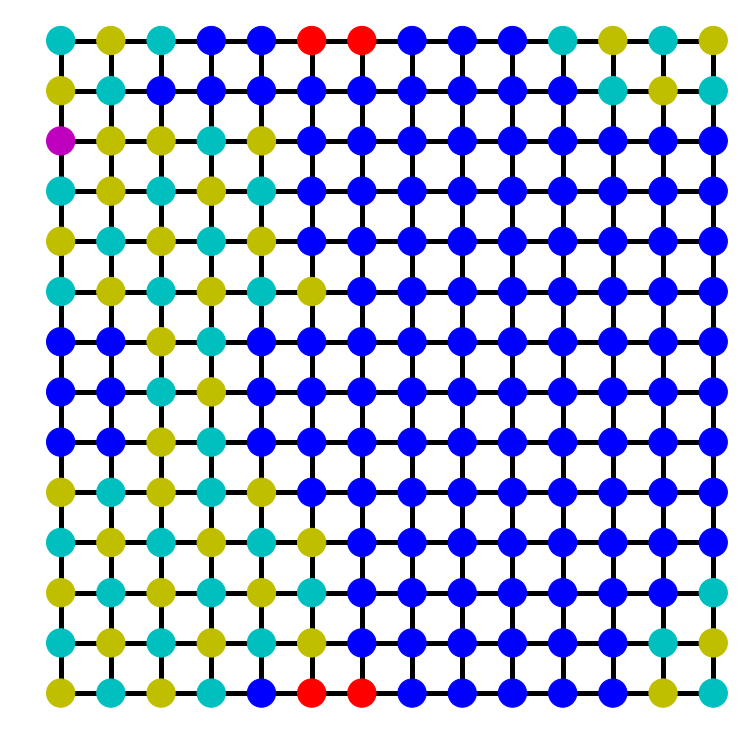}
    \end{subfigure}
     \caption{Snapshot of a (left) triangular and (right) square lattice for $q \approx q_c$. The colours encode the six possible translation-tables. The repeating patterns on the square lattice are \textit{stable misunderstandings.}}
     \label{fig:snap}
\end{figure}

\begin{figure}
    \centering
    \includegraphics[width=\textwidth]{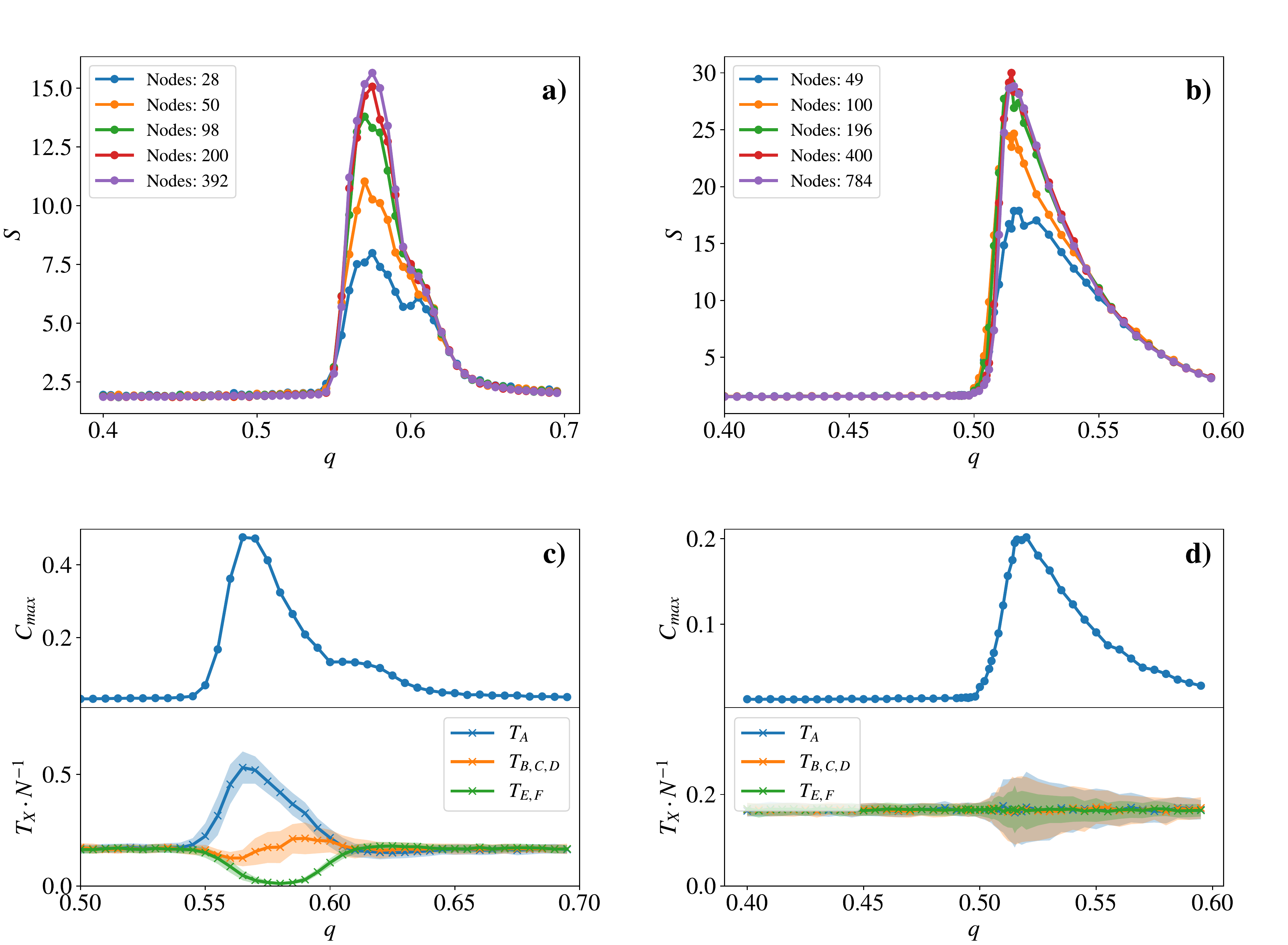}
\caption{
Variation of the mean cluster size ($S$) vs. the threshold parameter $q$ for different numbers of nodes in a \textbf{a)}~triangular and \textbf{b)}~square lattice (Simulated steps $t=20,000$, averaged over 100 runs). For $q \gg q_c$ the typical sizes are much smaller than the system size and the cluster size does not depend on $N$. For $q \approx q_c$ finite-size scaling appears and the cluster size depends on the system size. The similar behaviour of the two largest networks indicates that these systems did not finish growth after the $20,000$ time steps.\\
\textbf{c)}/\textbf{d)} (upper)~Relative cluster size of the largest cluster $C_{max}$ vs. the threshold parameter $q$ on an triangular ($N=392$)  and square ($N = 784$) lattice (After $t=20,000$ simulated steps, averaged over 100 runs). (lower)~The fraction of the different tables depending on $q$. Here, $T_X$ denotes the different type of possible translation tables as described in the full-text. The light area around the curves indicates the standard deviation. (After $t=10,000$ simulated steps, averaged over 100 runs)}
\label{fig:triang_grid_q}
\end{figure}

To gain more insight into the critical behaviour of our polycontextural network model, in the following we focus on regular triangular and square lattices. Fig.~\ref{fig:snap} shows a snapshot of two critical systems after the evolution of $t=20,000$ steps. Here, the colours of the nodes do not indicate their current colour but are illustrations for their respective translation table. Depending on the topology of the network, the clusters are only formed by equal translation tables or also by patterns of two alternating tables, indicating the occurrence of stable misunderstandings. This is in line with the analysis of the motifs in Sec.~\ref{sec:model}. 

Fig.~\ref{fig:triang_grid_q}~a) and~b) shows the mean cluster size over the value of $q$ for different system sizes for both the triangular and square lattices. As expected, for small values of $q$ the mean cluster size is close to one: the fluctuations in the system do not permit the buildup of correlations. At a rather sharp value $q = q_c$ we observe a sudden jump in the cluster size, indicating a phase transition. Then, for values $q > q_c$ the cluster size slowly decreases and converges to the initial mean cluster size of one. It is important to note that for $q \gg q_c$ the typical size of clusters is much smaller than the system size and hence the mean cluster size is not limited by the system size.

As already for the random regular graphs, a key quantity for our system is the size of the largest cluster. Figure~\ref{fig:triang_grid_q}~c) and~d) shows exemplary behaviours of the relative size of the largest cluster. For a small window of $q$ the largest cluster of the system spans the full system, indicating system-wide correlations between the tables. Additionally, the lower plots show the fraction of the six different translation tables, where $T_A$ denotes the identity, $T_{\{B,C,D\}}$ denote the three possible 2-cycle matrices and $T_{\{E,F\}}$ denote the two 3-cycle matrices, with $T_E \times T_F = 1$. As already observed in Fig.~\ref{fig:snap} the topology of the underlying network determines which translation tables cluster: At criticality, the triangular network predominantly consists of nodes that hold the identity matrix $T_A$. Contrarily, in the grid network, all translation tables have the same probability. However, as the diverging variance at the critical point already indicates, this is just an averaging effect. In a single system, only one type of cluster configuration wins, but the probability to win is equal for all configurations.

We will now turn to an analysis of the scale-freeness of the cluster size distribution.

\subsection{Self-similarity}

\begin{figure}
\centering
\includegraphics[width=\textwidth]{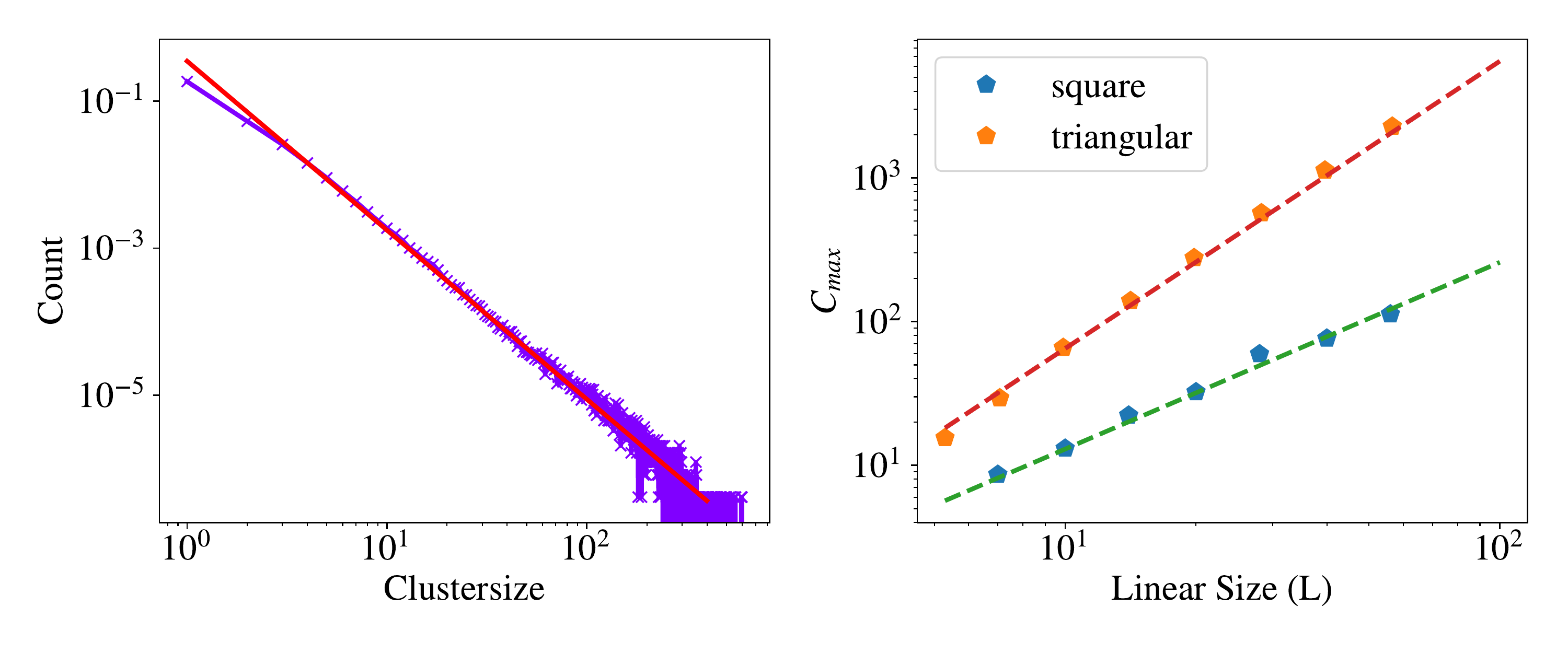}
\caption{
(left)~Cluster size distribution of the triangular network for $q=0.57$ at $t=5000$ ($N$=2450 nodes, averaged over $1000$ runs). The red line indicates a power law with exponent $\alpha = -2.3$.
(right)~Size of the largest cluster $C_{max}$ over the linear size $L = N^{0.5}$ of the system. Both systems were simulated with a $q \approx q_c$ for $t=20,000$ time steps. The dashed lines indicate the respective power-law fits with (green) $d_f = 1.3$ and (red) $d_f = 2$.
}
\label{fig:cluster_dimension}
\end{figure}

In some regard, the dynamics of our model are similar to grain growth and coarsening processes as observed in crystallisation. The clusters with compatible tables in our model (local consensus) play the role of domains with similar orientations in the coarsening process. A successful model to simulate such crystallisation processes is the Monte Carlo Potts Model~\cite{zollnerNormalGrainGrowth2008}. The Potts model is defined on a spatial grid where each grid point (often called Monte Carlo Unit) can be in one of $Q$ possible states that describe the respective orientation. At each time step, a random grid point is selected. Then, similar to the selection of a new translation table in our model, a new orientation is assigned to the selected grid point. However, in the Potts model, this assignment is on probation. The new orientation is selected with a probability that depends on the energy difference between the old and new state (with regard to the interactions with the nearest neighbours) as well as on an external parameter, the temperature $T$. It is well known that the cluster sizes in grain growth and particle coarsening show self-similarity~\cite{zollnerNormalGrainGrowth2008,mullinsStatisticalSelfSimilarity1986}. This means that during coarsening different system configurations reveal similar behaviour when scaled to the same scale: they are scale-invariant. Whether such self-similarity is also present in social dynamics is an ongoing debate~\cite{abbottPartSelfSimilaritySocial2010}. Due to the similarities between our model and the Potts model it is hence natural and interesting to ask if we also observe a scale-invariant behaviour in our model. A well-known characteristic of self-similarity is a power-law distribution of observable quantities. In the case of our model, this could e.g. be the cluster size distribution. In Fig.~\ref{fig:cluster_dimension}~(left) we show this distribution for the triangular network at $q = 0.57$, slightly larger than the critical value $q_c = 0.56$. In a log-log plot, this distribution shows a linear behaviour with slope $m = -2.3$, indicating a power law with an exponent of $\alpha = -2.3$. This cluster size distribution is hence scale-invariant. In terms of social systems, this would mean that an opinion structure that is found in small communities is equal to the structure that is found in large systems of interconnected communities. 

Self-similarity of the cluster size distribution would also imply that the size of the largest cluster $C_{\max}$ scales with the linear system size $L$ according to:
\begin{equation}
    C_{\max}(L) \sim L^{d_f}
\end{equation}
where $d_f$ is the (possibly) fractal dimension of the cluster~\cite{lesneScaleInvariance2012,tsakirisPercolationRandomlyDistributed2010}. In Fig.~\ref{fig:cluster_dimension}~(right) we observe the expected scaling for both lattices analysed and obtain the corresponding values for $d_f$. The quadratic lattice growth with a fractal dimension of $1.3$ indicating a rough boundary, whereas the triangular lattice growth with a fractal dimension of $2$, equal to the spatial dimension of the lattice, hence there is a high `surface tension' and the clusters are more compact with a smooth boundary. One should note that this determination of $d_f$ is not very accurate and one would need another approach to obtain a more exact version. This is, however, out of the scope of this manuscript and will be left for an upcoming publication.

  \begin{figure}
    \centering
    \includegraphics[width=.5\textwidth]{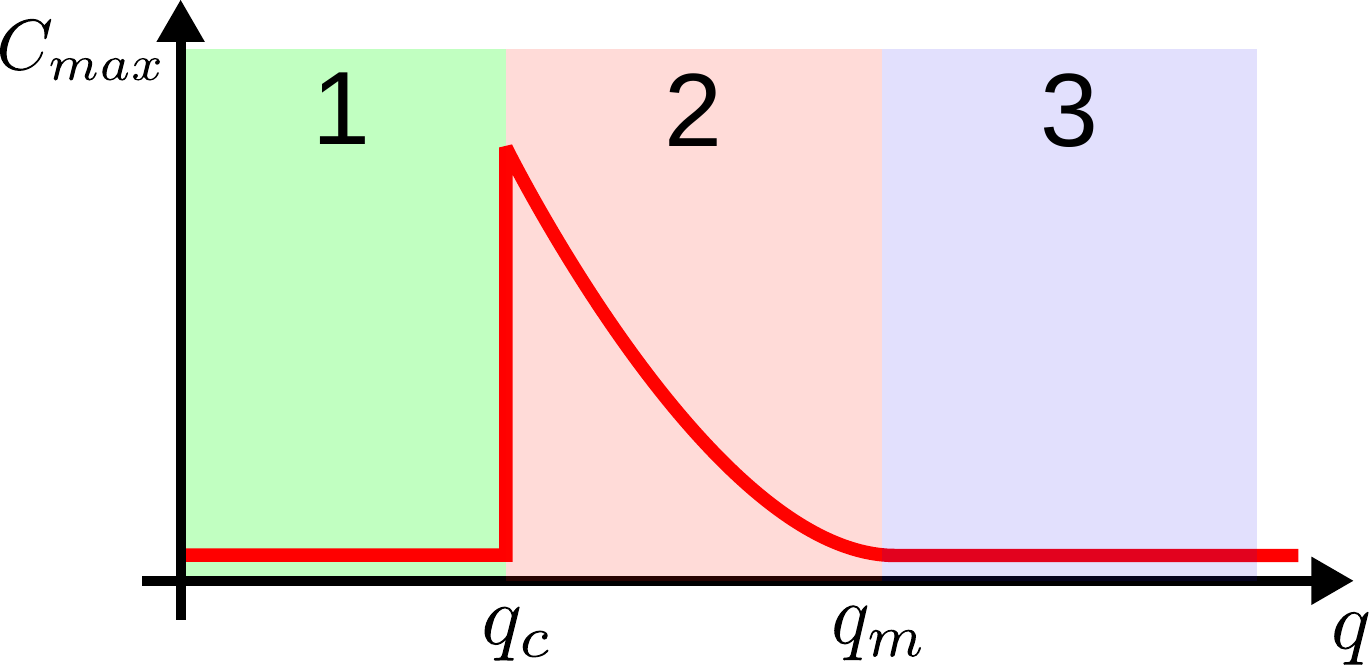}
    \caption{Phenomenological separation of three different phases. In phase 1 there is too much activity in the system. In phase 3 there is too little activity. Clusters can only grow in phase 2.}
    \label{fig:mechanistic}
\end{figure}   

With the results obtained, we are now in the position to explain mechanistically how clusters in our system are formed and how the observed self-similarity can be explained. Hereto, we draw on Fig.~\ref{fig:mechanistic}. If $q < q_c$ the system is in phase 1. There are too many fluctuations in the system such that no clusters of shared tables can emerge and possibly existing clusters are destroyed. For $q > q_m$ (phase 3) there is too little activity. Every single node behaves as a single nucleus of a new cluster and does not adapt to join other, possibly larger clusters. Upon a decrease of $q$ below $q_m$ neighbouring agents begin to form clusters. However, small nuclei of possibly incompatible clusters appear all over the system and grow (with dimension $d_f$) until they reach the boundary of other clusters. The result is a cluster–cluster competition between different incompatible clusters as it is also known for models like the naming game~\cite{baronchelliEmergenceConsensusPrimer2018}. The smaller $q$ the smaller is the probability that an initial nucleus appears. Close to $q$ with $q > q_c$ there is only a very small probability for an initial nucleus, but once a first nucleus is stabilised it can grow over the full system without being limited by another growing cluster.

\section{Discussion}
\label{sec:discussion}
We have shown that our simple model, the polycontextural network, has a phase where the world-views cluster globally, leading to a shared perception of signals, or global stable misunderstandings. In the following, we show how our model fits into the landscape of established models and discuss its implications.

\subsection{Clustering of Opinions}

The clustering of opinions within human populations is an important and ongoing research topic~\cite{masIndividualizationDrivingForce2010}. Based on empirically validated mechanisms like `homophily', different models tried to explain how opinion clustering might happen~\cite{dandekarBiasedAssimilationHomophily2013}. Most of these models are able to show a clustering of opinions. Depending on the detailed mechanisms of the model, adding noise can facilitate mono culture or maintains pluralism ~\cite{masIndividualizationDrivingForce2010,sirbuAlgorithmicBiasAmplifies2019}. A prominent source of such noise is the misinterpretation of information. Starting from an initial configuration where no predefined definition of a right or wrong interpretation of signals is given, our polycontextural network shows the build-up of a shared perception of signals. Our model hence provides a framework to understand the basic mechanisms of how a first basic understanding of information might emerge. A prerequisite most current models of social sciences (implicitly) rely on. 

\subsection{Polycontextural Logic}

\begin{figure}
    \centering
    \includegraphics{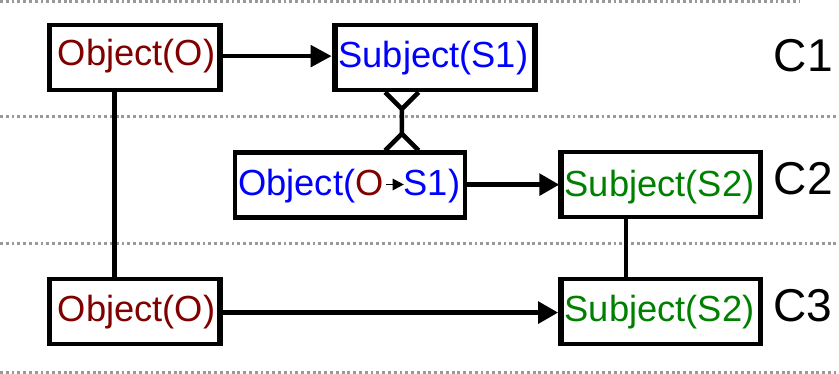}
    \caption{Sketch of the proemial relation. In each contexture (C1, C2, C3) a subject observes an object as indicated by the arrows (The direction of the arrows indicates the flow of information; Object $\rightarrow$ Subject reads: Subject observes Object). Figure and caption taken from~\cite{falkPhysicsOrganizedTransformations2021}.}
    \label{fig:proemial}
\end{figure}

Binary logic is a cornerstone of western thought and technology and hence an important component in our decision strategies and opinion formation processes. As a consequence, social dynamics around opinion formation are challenged by situations where facts can not easily be mapped to global and objective `true' and `false', as is e.g. the case in modern social phenomena, like the emergence and prevalence of fake news or the counter-phenomenon of fact-checking entities: \textit{What is obvious to us may be a lie from hell to our neighbour}~\cite{sireNamingElephantWorldview2015}. When facts are not perceived equally by two distinct observers, each of whom adheres to binary logic, we are in a situation, which the philosopher Gotthard Günther attempted to capture with his theory of polycontexturality\cite{guntherIdeeUndGrundriss1991}. He argues that each of the observers is applying the framework of Aristotelian logic consistently within this observer's own realm of observation, called \textit{contexture}. Each contexture has its own set of factual embedding of true and false. This alignment of their own subjective understanding with the community requires that each individual is able to self-reflect their own understanding. Based on Hegel's dialectics, Günther formally analysed how living beings with only subjective perception can interact and how they can become aware of their own subjectivity~\cite{guntherIdeeUndGrundriss1991,falkPhysicsOrganizedTransformations2021}. He assumed that every ordered combination of an observer (a subject) and the observed object form a contexture that each has its own classical two-valued logic. He found the mutual interaction between \textit{three} contextures to be a crucial requirement for successful communication. The three contextures are arranged in a structure as shown in Fig.~\ref{fig:proemial}, which Günther termed \textit{proemial relation}. In the first contexture (C1) an object (O) is observed by the subject S1. This contexture can become the observed object of a second subject (S2). Thereby, S2  observes the object of the first contexture \textit{as observed} by the observer S1 from the first contexture~\cite{klagenfurtTechnologischeZivilisationUnd2016}. Subsequently, within a third contexture S2 can compare the original object with its subjectified version. The proemial relation hence allows the single observers to reflect their own understandings of the world.

The dynamics of our polycontextural network can be interpreted in this manner: Let us assume two nodes A and B. The colour (the fact) of node A corresponds to the object O in the first contexture. This colour can subjectively be observed by node B (S1), which colours itself according to the result of this observation. Now, in the second contexture, node A (S2) can observe the colour of B (S1). Last (the third contexture), node A compares its own colour (O) with node B's colour (as observed by A itself) which is (based on A's standpoint) the own colour through the eyes of another~\cite{gunther1964bewusstsein}. Following this dynamics, node A is able to notice a possible misalignment between its own and node B's world-view (translation table). Our results hence indicate how the subjective observation of observers enables a self-reflection that can lead to the emergence of shared signals and provides a numerical example of Günther's and Hegel's philosophy.

Within social systems, communication and influence often lead to each of the social entities gathering `followers', supporters of their particular interpretation of a given set of facts and, hence, of their respective contexture. Contextures in this way become entrenched in society. Similarly, in our model, we observe the growth of clusters of similar world-views.

\subsection{Stable Misunderstandings}
Misunderstandings are a commonplace in communications~\cite{edwardsThatNotWhat2017}. Often, these misunderstandings arise from different interpretations of a message~\cite{robinson2001new,millerBlissfullyHappyReady2016}, whereby the interpretations are influenced by factors like personality and values~\cite{edwardsListeningMessageInterpretation2011}. Within our model, we observe the situation of \textit{stable misunderstandings}, where two agents with different world-views interpret messages differently but in a mutually compatible form. One might argue that such a permanent misunderstanding is an artificial situation and just an artefact of our model's design. However, misunderstandings are often subtle, persistent and difficult to uncover~\cite{SymposiumPreventiveSocial1958}. In this regard, a nice example are students' misunderstandings of quantum mechanical wave functions~\cite{porterGraduateStudentMisunderstandings2019}. Our model can be understood as a minimal model of such misunderstandings.

\section{Conclusion}
\label{sec:conclusion}
In this manuscript, we presented a new model to explain if and how consensus can appear between agents that can only judge based on a subjective understanding of the world. Focusing on two regular lattices, the triangular and the quadratic lattice, as well as random regular graphs we observed the emergence of a system-wide (and then by definition objective) understanding of signs within the network. This emergence depends on a single parameter that controls the volatility: If the agents are too volatile they change their convictions too often to form clusters of shared understandings. If the agents are not volatile enough they do not adapt to majority opinions and locally separated clusters of different convictions appear. Both phases are connected by a phase transition and only at the transition point the growth of a spanning cluster is possible. 

The findings of our model add to several ongoing discussions in social science as well as philosophy and computer science. 
Obviously, the study of our model on different types of networks is not exhausted. In this manuscript, we have restricted our analysis to the regular triangular and square grid lattices which already showed -- especially in terms of the cluster composition -- two quite different behaviours. The next natural step could hence be to observe how fast a consensus can be reached on random networks like ER or BA graphs. Here, it might also be interesting to introduce a degree-dependent threshold parameter such that nodes with a high degree are more convinced of their position and change their tables rarer. Additionally, it could be worthwhile to transform the dynamics of the model to a mechanism comparable to the $q$-voter model where each node observes $q$ neighbours and only changes its table if a given fraction of these neighbours have a wrong colour. To model people's behaviour a little more realistically an interesting modification of the model would also be to reduce the memory of the agents~\cite{shohamEmergenceSocialConventions1997}. Instead of the possibly infinite memory, one could restrict the agents to only remember the last $X$ observations. This would create more fluctuations and could avoid the creation of small but stable minority clusters. Additionally, it is interesting to increase the number of possible colours $C$ and thereby the number $C!$ of different world-views. This leads to a larger variety of possible misunderstandings and the possibility of stable and partially compatible world-views.

\bibliography{references}

\section*{Author contributions statement}

E.E. initiated the project. J.F., M-T.H. developed the model, J.F. ran the simulations and analysed the data. J.F. and M-T.H. wrote the manuscript. E.E., K.W. and M-T.H. supervised the project. All authors discussed the results and implications and commented on the manuscript at all stages.

\section*{Additional information}

On behalf of all authors, the corresponding author states that there is no conflict of interest.

\end{document}